\documentstyle[aps,prl,twocolumn,epsfig]{revtex}

\begin{document}
\title{Entanglement purification of Gaussian continuous variable quantum states}
\author{Lu-Ming Duan$^{1,2}$\thanks{%
Email: luming.duan@uibk.ac.at}, G. Giedke$^1$, J. I. Cirac$^1$, and P. Zoller%
$^1$}
\address{$^{1}$Institut f\"{u}r Theoretische Physik, Universit\"{a}t Innsbruck,\\
A-6020 Innsbruck, Austria \\
$^{2}$Laboratory of Quantum Communication and Quantum Computation,
University of Science and \\
Technology of China, Hefei 230026, China}
\maketitle

\begin{abstract}
We describe an entanglement purification protocol to generate maximally
entangled states with high efficiencies from two-mode squeezed states or
from mixed Gaussian continuous entangled states. The protocol relies on a
local quantum non-demolition measurement of the total excitation number of
several continuous variable entangled pairs. We propose an optical scheme to
do this kind of measurement using cavity enhanced cross--Kerr interactions.

\end{abstract}

Quantum communication, such as quantum key distribution and quantum
teleportation, is hampered by the difficulty to generate maximally entangled
states between distant nodes [1]. Due to loss and decoherence, in reality we
can only generate partially entangled states between distant sides [2].
Entanglement purification techniques are needed to concentrate maximally
entangled states from partially entangled states [3,4]. For qubit systems,
efficient entanglement purification protocols have been found [3-5]. But
none of these purification schemes have been realized experimentally due to
the great difficulty of performing repeated collective operations in
realistic quantum communication systems. Thus, it is of interest to consider
purification of continuous variable entanglement. The nonlocal Gaussian
continuous variable entangled states (i.e., states whose Wigner functions
are Gaussians) can be easily generated by transmitting two-mode squeezed
light, and this kind of entanglement has been demonstrated in the recent
experiment of continuous variable teleportation [6]. As the first choice for
performing continuous entanglement purification, one would consider direct
extensions of the purification schemes for qubit systems. But till now in
these extensions, no entanglement increase has been found for Gaussian
continuous entangled states [7]. Thus, the discussion should be extended to
a larger class of operations to purify continuous entangled states.
Braunstein et al. \cite{8} have proposed a simple error correction scheme
for continuous variables. However, it is not clear whether it can be used
for purification. In [9] a protocol to increase the entanglement for the
special case of pure two-mode squeezed states has been proposed, which is
based on conditional photon number subtraction; the efficiency, however,
seems to be an obstacle for its practical realization.

In this paper, we present an entanglement purification scheme with the
following properties: (i) For pure states it reaches the maximal allowed
efficiency in the asymptotic limit (when the number of pairs of modes goes
to infinity); (ii) It can be readily extended to distill maximally entangled
states from a relevant class of mixed Gaussian states which result from
losses in the light transmission. Furthermore, we propose and analyze a
scheme to implement this protocol experimentally using high finesse cavities
and cross--Kerr nonlinearities. Our purification protocol generates
maximally entangled states in finite dimensional Hilbert spaces. The
entanglement in the continuous partially entangled state is transformed to
the maximally entangled state with a high efficiency. We begin the paper by
describing the entanglement purification protocol for pure two mode squeezed
states, and then extend the protocol to include mixed Gaussian continuous
states, and last describe the physical implementation of the purification
protocol.

First assume that we have generated $m$ entangled pairs $A_{i},B_{i}$ $%
\left( i=1,2,\cdots m\right) $ between two distant sides A and B. Each pair
of modes $A_{i},B_{i}$ are prepared in the two mode squeezed state $\left|
\Psi \right\rangle _{A_{i}B_{i}}$, which in the number basis has the form 
\begin{equation}
\left| \Psi \right\rangle _{A_{i}B_{i}}=\sqrt{1-\lambda ^{2}}\stackrel{%
\infty }{%
\mathrel{\mathop{\sum }\limits_{n=0}}%
}\lambda ^{n}\left| n\right\rangle _{A_{i}}\left| n\right\rangle _{B_{i}},
\end{equation}
where $\lambda =\tanh \left( r\right) ,$ and $r$ is the squeezing parameter 
\cite{10}. For and only for a pure state, the entanglement is uniquely
quantified by the von Neumann entropy of the reduced density operator of its
one-component. The entanglement of the state (1) is thus given by $E\left(
\left| \Psi \right\rangle _{A_{i}B_{i}}\right) =\cosh ^{2}\left( r\right)
\log \left( \cosh ^{2}\left( r\right) \right) -\sinh ^{2}\left( r\right)
\log \left( \sinh ^{2}\left( r\right) \right) $. The joint state $\left|
\Psi \right\rangle _{\left( A_{i}B_{i}\right\} }$ of the $m$ entangled pairs
is simply the product of all the $\left| \Psi \right\rangle _{A_{i}B_{i}}$,
which can be rewritten as 
\begin{equation}
\left| \Psi \right\rangle _{\left( A_{i}B_{i}\right\} }=\left( 1-\lambda
^{2}\right) ^{\frac{m}{2}}\stackrel{\infty }{%
\mathrel{\mathop{\sum }\limits_{j=0}}%
}\lambda ^{j}\sqrt{f_{j}^{\left( m\right) }}\left| j\right\rangle _{\left(
A_{i}B_{i}\right\} },
\end{equation}
where $\left( A_{i}B_{i}\right\} $ is abbreviation of the symbol $A_{1},B_{1}
$, $A_{2},B_{2},$ $\cdots $ and $A_{m},B_{m}$, and the normalized state $%
\left| j\right\rangle _{\left( A_{i}B_{i}\right\} }$ is defined as 
\begin{eqnarray}
\left| j\right\rangle _{\left( A_{i}B_{i}\right\} } &=&\frac{1}{\sqrt{%
f_{j}^{\left( m\right) }}}\stackrel{i_{1}+i_{2}+\cdots +i_{m}=j}{%
\mathrel{\mathop{\sum }\limits_{i_{1},i_{2},\cdots ,i_{m}}}%
}  \nonumber \\
&&\left| i_{1},i_{2},\cdots ,i_{m}\right\rangle _{\left( A_{i}\right\}
}\otimes \left| i_{1},i_{2},\cdots ,i_{m}\right\rangle _{\left(
B_{i}\right\} }.
\end{eqnarray}
The function $f_{j}^{\left( m\right) }$ in Eq. (2) and (3) is given by $%
f_{j}^{\left( m\right) }=\frac{\left( j+m-1\right) !}{j!\left( m-1\right) !}.
$ To concentrate entanglement of these $m$ entangled pairs, we perform a QND
measurement of the total excitation number $n_{A_{1}}+n_{A_{2}}+\cdots
+n_{A_{m}}$ on the A side (we will describe later how to implement this
measurement experimentally). The QND measurement projects the state $\left|
\Psi \right\rangle _{\left( A_{i}B_{i}\right\} }$ onto a two-party maximally
entangled state $\left| j\right\rangle _{\left( A_{i}B_{i}\right\} }$ with
probability $p_{j}=\left( 1-\lambda ^{2}\right) ^{m}\lambda
^{2j}f_{j}^{\left( m\right) }.$The entanglement of the outcome state $\left|
j\right\rangle _{\left( A_{i}B_{i}\right\} }$ is given by $E\left( \left|
j\right\rangle _{\left( A_{i}B_{i}\right\} }\right) =\log \left(
f_{j}^{\left( m\right) }\right) $. The quantity $\Gamma _{j}=E\left( \left|
j\right\rangle _{\left( A_{i}B_{i}\right\} }\right) /E\left( \left| \Psi
\right\rangle _{A_{i}B_{i}}\right) $ defines the entanglement increase
ratio, and if $\Gamma _{j}>1$, we get a more entangled state. Even with a
small number $m$, the probability getting a more entangled state is quite
high. It can be easily proven that if $m$ goes to infinity, with unit
probability we would get a maximally entangled state with entanglement $%
mE\left( \left| \Psi \right\rangle _{A_{i}B_{i}}\right) .$ This ensures that
this method is optimal in this limit, analogous to the purification protocol
presented in [3] for the qubit case. For any finite number of entangled
pairs, the present purification protocol is more efficient than that in [3],
since it takes advantage of the special relations between the coefficients
in the two-mode squeezed state.

An interesting feature of this entanglement purification protocol is that
for any measurement outcome $j\neq 0$, we always get a useful maximally
entangled state in some finite Hilbert space, though the entanglement of the
outcome state $\left| j\right\rangle _{\left( A_{i}B_{i}\right\} }$ does not
necessarily exceed that of the original state $\left| \Psi \right\rangle
_{A_{i}B_{i}}$ if $j$ is small. It is also interesting to note that a small
alternation of this scheme provides a useful method for preparing GHZ--like
states in high dimensional Hilbert spaces \cite{11}. The key point is that
the modes $B_{i}$ need not be at the same side in the protocol. Assume we
have two entangled pairs $B,A_{1}$ and $A_{2},C$ distributed at three sides
B, A, C, with each pair being prepared in the state (1). Then a local QND
measurement of the modes $A_{1},A_{2}$ at the A side with the outcome $j\neq
0$ generates a three-party GHZ state in the $\left( j+1\right) $-dimensional
Hilbert space. Obviously, if we have $m$ entangled pairs, we can generate a $%
\left( m+1\right) $-party GHZ state using this method.

In reality, the light transmission will be unavoidably subjected to loss,
and then we will not start from an ideal two mode squeezed state, but
instead from a mixed state described by the following master equation 
\begin{equation}
\stackrel{.}{\rho }=-i\left( H_{\text{eff}}\rho -\rho H_{\text{eff}%
}^{\dagger }\right) +\stackrel{m}{%
\mathrel{\mathop{\sum }\limits_{i=1}}%
}\left( \eta _{A}a_{A_{i}}\rho a_{A_{i}}^{\dagger }+\eta _{B}a_{B_{i}}\rho
a_{B_{i}}^{\dagger }\right)
\end{equation}
where $\rho $ is the density operator of the $m$ entangled pairs with $\rho
\left( 0\right) =\left| \Psi \right\rangle _{\left( A_{i}B_{i}\right\}
}\left\langle \Psi \right| $, the ideal two mode squeezed state, and the
effective Hamiltonian 
\begin{equation}
H_{\text{eff}}=-i\stackrel{m}{%
\mathrel{\mathop{\sum }\limits_{i=1}}%
}\left( \frac{\eta _{A}}{2}a_{A_{i}}^{\dagger }a_{A_{i}}+\frac{\eta _{B}}{2}%
a_{B_{i}}^{\dagger }a_{B_{i}}\right) .
\end{equation}
In Eqs. (4) and (5), $a_{\alpha _{i}}$ denotes the annihilation operator of
the mode $\alpha _{i}$ $\left( \alpha =A\text{ or }B\right) $, and we have
assumed that the damping rates $\eta _{A}$ and $\eta _{B}$ are the same for
all the $m$ entangled pairs based on symmetry considerations, but $\eta _{A}$
and $\eta _{B}$ may be different to each other.

In many practical cases, it is reasonable to assume that the light
transmission noise is small. Let $\tau $ denote the transmission time, then $%
\eta _{A}\tau $ and $\eta _{B}\tau $ are small factors. In the language of
quantum trajectories \cite{10}, to the first order of $\eta _{A}\tau $ and $%
\eta _{B}\tau $, the final state of the $m$ entangled pairs is either $%
\left| \Psi ^{\left( 0\right) }\right\rangle _{\left( A_{i}B_{i}\right\}
}\propto e^{-iH_{\text{eff}}\tau }\left| \Psi \right\rangle _{\left(
A_{i}B_{i}\right\} }$ with no quantum jumps occurred, or $\left| \Psi
^{\left( \alpha _{i}\right) }\right\rangle _{\left( A_{i}B_{i}\right\}
}\propto \sqrt{\eta _{\alpha }\tau }a_{\alpha _{i}}\left| \Psi \right\rangle
_{\left( A_{i}B_{i}\right\} }$ with a jump occurred in the $\alpha _{i}$
channel $\left( \alpha =A,B\text{ and }i=1,2,\cdots ,m\right) $. The final
density operator is a mixture of all these possible states. To purify
entanglement from the mixed state, we perform QND measurements of the total
excitation number on both sides A and B, and the measurement results are
denoted by $j_{A}$ and $j_{B}$, respectively. We then compare $j_{A}$ and $%
j_{B}$ through classical communication, and keep the outcome state if and
only if $j_{A}=j_{B}$. Let $P_{A}^{\left( j\right) }$ and $P_{B}^{\left(
j\right) }$ denote the projections onto the eigenspaces of the corresponding
total number operators $\stackrel{m}{%
\mathrel{\mathop{\sum }\limits_{i=1}}%
}a_{A_{i}}^{\dagger }a_{A_{i}}$ and $\stackrel{m}{%
\mathrel{\mathop{\sum }\limits_{i=1}}%
}a_{B_{i}}^{\dagger }a_{B_{i}}$ with eigenvalue $j$, respectively. It is
easy to show that 
\begin{eqnarray}
P_{A}^{\left( j\right) }P_{B}^{\left( j\right) }\left| \Psi ^{\left(
0\right) }\right\rangle _{\left( A_{i}B_{i}\right\} } &=&\left|
j\right\rangle _{\left( A_{i}B_{i}\right\} },  \nonumber \\
P_{A}^{\left( j\right) }P_{B}^{\left( j\right) }\left| \Psi ^{\left( \alpha
_{i}\right) }\right\rangle _{\left( A_{i}B_{i}\right\} } &=&0.
\end{eqnarray}
So if $j_{A}=j_{B}=j$, the outcome state is the maximally entangled state $%
\left| j\right\rangle _{\left( A_{i}B_{i}\right\} }$ with entanglement $\log
\left( f_{j}^{\left( m\right) }\right) $. The probability to get the state $%
\left| j\right\rangle _{\left( A_{i}B_{i}\right\} }$ is now given by $%
p_{j}^{^{\prime }}=\left( 1-\lambda ^{2}\right) ^{m}\lambda
^{2j}f_{j}^{\left( m\right) }e^{-\left( \eta _{A}+\eta _{B}\right) \tau j}.$
It should be noted that the projection operators $P_{A}^{\left( j\right)
}P_{B}^{\left( j\right) }$ cannot eliminate the states obtained from the
initial state $\left| \Psi \right\rangle _{\left( A_{i}B_{i}\right\} }$ by a
quantum jump on each side A and B. The total probability for occur of this
kind of quantum jumps is proportional to $m^{2}\overline{n}^{2}\eta _{A}\eta
_{B}\tau ^{2}$. So the condition for small transmission noise requires $m^{2}%
\overline{n}^{2}\eta _{A}\eta _{B}\tau ^{2}\ll 1,$ where $\overline{n}=\sinh
^{2}\left( r\right) $ is the mean photon for a single mode.

In the purification for mixed entanglement, we need classical communication
(CC) to confirm that the measurement outcomes of the two sides are the same,
and during this CC, we implicitly assume that the storage noise for the
modes is negligible. In fact, that the storage noise is much smaller than
the transmission noise is a common assumption taken in all the entanglement
purification schemes which need the help of repeated CCs [4,5]. If we also
make this assumption for continuous variable systems, there exists another
simple configuration for the purification protocol to work. We put the
generation setup for two-mode squeezed states on the A side. After state
generation, we keep the modes $A_{i}$ on side A with a very small storage
loss rate $\eta _{A}$, and at the same time the modes $Bi$ are transmitted
to the distant side B with a loss rate $\eta _{B}\gg \eta _{A}.$ We call
this a configuration with an asymmetric transmission noise. In this
configuration, the purification protocol is exactly the same as that
described in the above paragraph. We note that the component in the final
mixed density operator which is kept by the projection $P_{A}^{\left(
j\right) }P_{B}^{\left( j\right) }$ should be subjected to the same times of
quantum jumps on each side A and B. We want this component to be a maximally
entangled state. This requires that the total probability for sides A and B
to subject to the same nonzero times of quantum jumps should be very small.
This total probability is always smaller than $\overline{n}\eta _{A}\tau $,
despite how large the damping rate $\eta _{B}$ is. So the working condition
of the purification protocol in the asymmetric transmission noise
configuration is given by $\overline{n}\eta _{A}\tau \ll 1$. The loss rate $%
\eta _{B}$ can be large. The probability to get the maximally entangled
state $\left| j\right\rangle _{\left( A_{i}B_{i}\right\} }$ is still given
by $p_{j}^{^{\prime }}=\left( 1-\lambda ^{2}\right) ^{m}\lambda
^{2j}f_{j}^{\left( m\right) }e^{-\left( \eta _{A}+\eta _{B}\right) \tau j}.$

For continuous variable systems the assumption of storage with a very small
loss rate is typically unrealistic. If this is the case, then we can use the
following simple method to circumvent the storage problem. Note that the
purpose to distill maximally entangled states is to directly apply them in
some quantum communication protocols, such as in quantum cryptography or in
quantum teleportation. So we can modify the above purification protocol by
the following procedure: right after the state generation, we take a QND
measurement of the total excitation number on side A and get a measurement
result $j_{A}$. Then we do not store the outcome state on side A, but
immediately use it (e.g., perform the corresponding measurement as required
by a quantum cryptography protocol \cite{12}). During this process, the
modes $B_{i}$ are being sent to the distant side B, and when they arrive, we
take another QND measurement of the total excitation number 
of the modes $B_{i}$ and get a outcome $j_{B}$.
The resulting state on side B can be directly used (for quantum cryptography
for instance) if $j_{A}=j_{B},$ and discarded otherwise. By this method, we
formally get maximally entangled states through posterior confirmation, and
at the same time we need not store the modes on both sides.

To experimentally implement the above purification scheme, we need first
generate Gaussian continuous entangled states between two distant sides, and
then perform a local QND measurement of the total excitation number of
several entangled pairs. Here we propose a promising experimental scheme,
which uses high finesse optical cavity to carry continuous entangled states
and cavity enhanced cross Kerr interactions to realize the local QND\
measurement. It is possible to generate Gaussian continuous entangled states
between two distant cavities \cite{13}. We can transmit and then couple the
two output lights of the nondegenerate optical parametric amplifier to
distant high finesse cavities. The steady state of the cavities is just a
Gaussian continuous entangled state described by the solution of Eq. (4)
after taking into account of the propagation loss [14]. The difficult part
is to perform a QND measurement of the total photon number contained in
several local cavities. We use the setup depicted in Fig. 1 to attain this
goal. (For convenience, we use the two-cavity measurement as an example to
illustrate the method. Extension of the measurement method to multi-cavity
cases is straightforward.)

\begin{figure}[tbp]
\epsfig{file=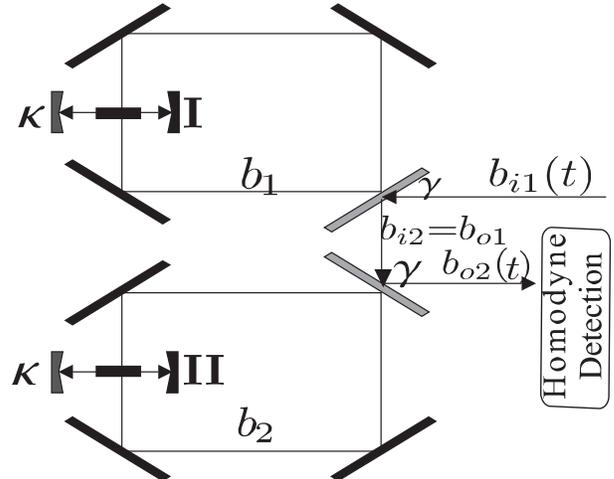,width=8cm}
\caption{Schematic experimental setup to measure the total photon number $%
n_{1}+n_{2}$ contained in the cavities I and II. The cavities I and II, each
with a small damping rate $\protect\kappa $ and with a cross Kerr medium
inside, are put respectively in a bigger ring cavity. The ring cavities with
the damping rate $\protect\gamma $ are used to enhance the cross Kerr
interactions. A strong cotinuous coherent driving light $b_{i1}\left(
t\right) $ is incident on the first ring cavity, whose output $b_{o1}$ is
directed to the second ring cavity. The output $b_{o2}\left( t\right) $ of
the second ring cavity is continuously observed through a homodyne
detection. }
\end{figure}

The measurement model depicted in Fig. 1 is an example of the cascaded
quantum system \cite{10}. The incident light $b_{i1}$ can be expressed as $%
b_{i1}=b_{i1}^{^{\prime }}+g\sqrt{\gamma },$ where $g\sqrt{\gamma }$ ($g$ is
a large dimensionless factor) is a constant driving field, and $%
b_{i1}^{^{\prime }}$ is the standard vacuum white noise, satisfying $%
\left\langle b_{i1}^{^{\prime }\dagger }\left( t\right) b_{i1}^{^{\prime
}}\left( t^{^{\prime }}\right) \right\rangle =0$ and $\left\langle
b_{i1}^{^{\prime }}\left( t\right) b_{i1}^{^{\prime }\dagger }\left(
t^{^{\prime }}\right) \right\rangle =\delta \left( t-t^{^{\prime }}\right) .$
The Hamiltonian for the Kerr medium is assumed to be $H_{i}=\hbar \chi
n_{i}b_{i}^{\dagger }b_{i},$ $\left( i=1\text{ or }2\right) ,$where $b_{i}$
is the annihilation operator for the ring cavity mode, and $\chi $ is the
cross-phase modulation coefficient. The self-phase modulation can be made
much smaller than the cross phase modulation with some resonance conditions
for the Kerr medium, and thus is negligible [15,16]. In the frame rotating
at the optical frequencies, the Langevin equations describing the dynamics
in the two ring cavities have the form 
\begin{eqnarray}
\stackrel{.}{b}_{1} &=&-i\chi n_{1}b_{1}-\frac{\gamma }{2}b_{1}-\sqrt{\gamma 
}b_{i1}^{^{\prime }}-g\gamma ,  \nonumber \\
\stackrel{.}{b}_{2} &=&-i\chi n_{2}b_{2}-\frac{\gamma }{2}b_{2}-\sqrt{\gamma 
}b_{i2},
\end{eqnarray}
with the boundary conditions (see Fig.\ 1) $b_{i2}=b_{o1}=b_{i1}^{^{\prime
}}+g\sqrt{\gamma }+\sqrt{\gamma }b_{1}$ and $b_{o2}=b_{i2}+\sqrt{\gamma }%
b_{2}.$ In the realistic case $\gamma \gg \chi \left\langle
n_{i}\right\rangle ,$ $\left( i=1,2\right) $, we can adiabatically eliminate
the cavity modes $b_{i}$, and express the final output $b_{o2}$ of the
second ring cavity as an operator function of the observable $n_{1}+n_{2}$.
The experimentally measured quantity is the integration of the homodyne
photon current over the measurement time $T.$ Choosing the phase of the
driving field so that $g=i\left| g\right| $, the measured observable
corresponds to the operator 
\begin{eqnarray}
X_{T} &=&\frac{1}{T}\int_{0}^{T}\frac{1}{\sqrt{2}}\left[ b_{o2}\left(
t\right) +b_{o2}^{\dagger }\left( t\right) \right] dt  \nonumber \\
&\approx &\frac{4\sqrt{2}\left| g\right| \chi }{\sqrt{\gamma }}\left(
n_{1}+n_{2}\right) +\frac{1}{\sqrt{T}}X_{T}^{\left( b\right) },
\end{eqnarray}
where $X_{T}^{\left( b\right) }=\frac{1}{\sqrt{2}}\left(
b_{T}+b_{T}^{\dagger }\right) $, and $b_{T}$, satisfying $\left[
b_{T},b_{T}^{\dagger }\right] =1$, is defined by $b_{T}=\frac{1}{\sqrt{T}}%
\int_{0}^{T}b_{i1}^{^{\prime }}\left( t\right) dt.$ Equation (8) assumes $%
\gamma \gg \chi \left\langle n_{i}\right\rangle $ and $e^{-\gamma T}\ll 1$.
There are two different contributions in Eq.\ (8). The first term represents
the signal, which is proportional to $n_{1}+n_{2}$, and the second term is
the vacuum noise. The distinguishability of this measurement is given by $%
\delta n=\frac{\sqrt{\gamma }}{8\left| g\right| \chi \sqrt{T}}.$ If $\delta
n<1$, i.e., if the measuring time $T>\frac{\gamma }{64\left| g\right|
^{2}\chi ^{2}},$ we effectively perform a measurement of $n_{1}+n_{2}$; and
if $T$ is also smaller than $\frac{1}{\kappa \left\langle n_{i}\right\rangle 
}$, the photon loss in the cavities I and II during the measurement is
negligible. So the setup gives an effective QND measurement of the total
photon number operator $n_{1}+n_{2}$ under the condition 
\begin{equation}
\frac{\gamma }{64\left| g\right| ^{2}\chi ^{2}}<T<\frac{1}{\kappa
\left\langle n_{i}\right\rangle }.
\end{equation}
This condition seems to be feasible with the present technology. For
example, if we assume the cross Kerr interaction is provided by the
resonantly enhanced Kerr nonlinearity as considered and demonstrated in
[15,16], the Kerr coefficient $\chi /2\pi \sim 0.1MHz$ would be obtainable
[17]. We can choose the decay rates $\kappa /2\pi \sim 4MHz$ and $\gamma
/2\pi \sim 100MHz,$ and let the dimensionless factor $g\sim 100$ (for a
cavity with cross area $S\sim 0.5\times 10^{-4}cm^{2}$, $g\sim 100$
corresponds a coherent driving light with intensity about $40mWcm^{-2}$).
The mean photon number $\left\langle n_{1}\right\rangle =\left\langle
n_{2}\right\rangle =\sinh ^{2}\left( r\right) \sim 1.4$ for a practical
squeezing parameter $r\sim 1.0.$ With the above parameters, Eq. (9) can be
easily satisfied if we choose the measuring time $T\sim 8ns$. More favorable
values for the parameters are certainly possible.

To bring the above proposal into a real experiment, there are several
imperfect effects which should be considered. These imperfections include
phase instability of the driving field, imbalance between the two ring
cavities, light absorption of the Kerr media and the mirrors, self phase
modulation effects, light transmission loss between the ring cavities, and
inefficiency of the detectors. To realize a QND measurement, the
imperfections should be small enough. We have deduced quantitative
requirements for all the imperfections listed above [18]. With the
parameters given in the above paragraph, all these requirements can be met
experimentally.

We thank P. Grangier and S. Parkins for discussions. This work was supported
by the Austrian Science Foundation, by the European TMR network Quantum
Information, and by the Institute for Quantum Information.


\begin{references}
\bibitem{1}  C. H. Bennett, Phys. Today 48 (10), 24 (1995)

\bibitem{2}  J. I. Cirac et al., Phys. Rev. Lett. 78, 3221 (1997); S. J.
Enk, J. I. Cirac, and P. Zoller, Science 279, 205 (1998).

\bibitem{3}  C. H. Bennett et al., Phys. Rev. A 53, 2046 (1996).

\bibitem{4}  C. H. Bennett et al., Phys. Rev. Lett. 76, 722 (1996).

\bibitem{5}  C. H. Bennett et al., Phys. Rev. A 54, 3824 (1996).

\bibitem{6}  A. Furusawa, et al., Science 282, 706 (1998).

\bibitem{7}  S. Parker, S. Bose, and M. B. Plenio, e-print quant-ph/9906098.

\bibitem{8}  S. L. Braunstein, Nature 394, 47 (1998); Phys. Rev. Lett. 80,
4084 (1998); S. Lloyd and J. J.-E Slotine, Phys. Rev. Lett. 80, 4088 (1998).

\bibitem{9}  T. Opatrny, G. Kurizki, and D.-G., Welsch, e-print
quant-ph/9907048.

\bibitem{10}  C. W. Gardiner and P. Zoller, {\it Quantum Noise}
(Springer--Verlag, Berlin, 1999).

\bibitem{11}  D. Greenberger et al., Am. J. Phys. 58, 1131 (1990); J. W. Pan
et al., Nature (to be published); G. M. D'Ariano et al. e-print
quant-ph/9906067.

\bibitem{12}  M. Hillery, e-print quant-ph/9909006.

\bibitem{13}  N. Ph. Georgiades et al., Phys. Rev. Lett. 75, 3426 (1995).

\bibitem{14}  A. S. Parkins and H. J. Kimble, e-print quant-ph/9907049.

\bibitem{15}  A. Imamoglu at al., Phys. Rev. Lett. 79, 1467 (1997); 81, 2836
(1998).

\bibitem{16}  L. V. Hau, et al., Nature 397, 594 (1999).

\bibitem{17}  In fact, Ref. [15] considered a configuation, yielding a Kerr
coefficient $\chi \sim 100MHz$, to realize a single-photon turnstile device.
But the estimation there puts a stringent limit on the required cavity
parameters [K. M. Gheri {\em et al.}, Phys. Rev. A 60, R2673, 1999]. We take
a much more moderate estimation of the relavent parameters and find $\chi
/2\pi \sim 0.1MHz$ is obtainable. This value of the Kerr coefficient is
large enough for performing the QND measurement.

\bibitem{18}  L. M. Duan et. al., (to be published).
\end{references}
\end{document}